\documentclass[final,12pt]{elsarticle}

\usepackage{graphicx}

\usepackage{amssymb}
\usepackage{amsmath}

\usepackage{booktabs}
\usepackage{array}

\DeclareMathOperator{\cov}{cov} 

\newcolumntype{C}{>{$}c<{$}}
\newcolumntype{F}{>{$}m{0.2 \columnwidth}<{$}}

\journal{Physica A}

\begin{document}

\begin{frontmatter}



\title{A Bayesian Networks Approach to Operational Risk}


\author[carma,formit]{V. Aquaro}

\author[uniba,infn_ba]{M. Bardoscia \corref{mail}}
\cortext[mail]{Corresponding author mail: \texttt{marco.bardoscia@ba.infn.it} and tel: \texttt{+390805442178}}

\author[uniba,infn_ba]{R. Bellotti}

\author[carma]{A. Consiglio}

\author[infn_ba]{\\F. De Carlo}

\author[eco]{G. Ferri}

\address[carma]{CARMA, Research Consortium for Risk Management Automation, \\via Mitolo 17B, I-70124 Bari, Italy}
\address[formit]{Formit Servizi S.p.A., via C.Conti Rossini 26, I-00147 Roma, Italy}
\address[uniba]{Dipartimento Interateneo di Fisica ``\emph{M.Merlin}'', Universit\`a degli Studi di Bari e Politecnico di Bari, via Amendola 173, I-70126 Bari, Italy}
\address[infn_ba]{Istituto Nazionale di Fisica Nucleare, Sezione di Bari, \\ via Amendola 173, I-70126 Bari, Italy}
\address[eco]{Dipartimento di Scienze Economiche e Metodi Matematici, Universit\`a degli Studi di Bari, via C.Rosalba 53, I-70124, Italy}

\begin{abstract}
A system for Operational Risk management based on the computational paradigm of Bayesian Networks is presented. The algorithm allows the construction of a Bayesian Network targeted for each bank and takes into account in a simple and realistic way the correlations among different processes of the bank. The internal losses are averaged over a variable time horizon, so that the correlations at different times are removed, while the correlations at the same time are kept: the averaged losses are thus suitable to perform the learning of the network topology and parameters; since the main aim is to understand the role of the correlations among the losses, the assessments of domain experts are not used. The algorithm has been validated on synthetic time series. It should be stressed that the proposed algorithm has been thought for the practical implementation in a mid or small sized bank, since it has a small impact on the organizational structure of a bank and requires an investment in human resources which is limited to the computational area.
\end{abstract}

\begin{keyword}
Operational Risk \sep Complex Systems \sep Bayesian Networks \sep Time Series \sep Value-at-Risk \sep Different-Times Correlations

\PACS 89.65.Gh \sep 05.45.Tp

\MSC 91B30 \sep 91B84 \sep 37M10 \sep 62M10

\end{keyword}

\end{frontmatter}



\section{Introduction} \label{sec:introduction}

In the past years a powerful set of tools to study complexity has been developed  by physicists and applied to economic and social systems; among the several topics under investigation the quantitative estimation and management of several typologies of risks \cite{mcneil-frey-embrechts}, like financial risk \cite{mantegna-stanley, bouchard-potters, mccauley, ccc, decarlo-et-al} and operational risk \cite{voit, khun-anand} has recently emerged.

\emph{Operational Risk} (OR) is defined as ``the risk of [money] loss resulting from inadequate or failed internal processes, people and systems or from external events'' \cite{basil}, including legal risk, but excluding strategic and reputation linked risks. Since it depends on a family of heterogeneous causes, in the past only few banks dealt with OR management. Starting from 2005 the approval of \emph{``The New Basel Capital Accord''} (Basel II) has substantially changed this picture: in fact OR is now considered a critical risk factor and banks are prescribed to cope with it setting aside a certain capital charge.

Basel II proposes three methods to determine this capital: i) the \emph{Basic Indicator Approach} sets it to $15\%$ of the bank's gross income; ii) the \emph{Standardized Approach} is a simple generalization of the Basic Indicator Approach: the percentage of the gross income is different for each Business Line and varies between $12\%$ and $18\%$; iii) the \emph{Advanced Measurement Approach} (AMA) allows each bank to use an internally developed procedure to estimate the impact of OR. Both the Basic Indicator Approach and the Standardized Approach seems overly simplistic, since in some way they suppose that the exposure of a bank to operational losses is proportional to its size. On the other hand, an AMA not only helps a bank to set aside the required capital charge, but may even allow the \emph{OR management}, in the prospect of limiting the amount of future losses.

Each AMA has to take into account two types of historical operational losses: the internal ones, collected by the bank itself, and the external ones which may belong to a database shared among several banks. Nevertheless, due to the recent interest for OR, only small and not adequately accurate historical databases exist and this is why each AMA is required to use also assessment data produced by experts. In addition, Basel II provides a classification of operational losses in $8$ Business Lines and $7$ Loss Event Types which has to be shared by all the AMAs. Finally, AMAs usually identify the capital charge with the Value-at-Risk (VaR) over the time horizon of $1$ year and with a confidence level of $99.9\%$, defined as the maximum potential loss not to be exceeded in $1$ year with confidence level of $99.9\%$, i.e.\ the $99.9$ percentile of the yearly loss distribution; this implies that the probability of registering a loss being less than or equal to the value of the VaR in $1$ year is equal to $0.999$ or, equivalently, that a loss larger than the value of the VaR may occur on average every $1000$ years.

Among the AMA methods, the most widely used is the \emph{Loss Distribution Approach} (LDA). In LDA the distribution of frequency and the distribution of impact (severity) modeling the operational losses are separately studied for each of the $56$ pairs $(\text{Business Line}, \text{Loss Event Type})$. LDA makes two crucial assumptions: i) frequency and severity distributions are independent for each pair; ii) the distributions of each pair are independent from the distributions of \emph{all the other} pairs. In other words LDA neglects the correlations possibly existing between the frequency or the severity of the losses occurring in different pairs.

The idea of exploiting BNs to study OR has already been proposed in Refs. \cite{cruz, alexander, neil-fenton-tailor, cowell-verral-yoon, bonafede-giudici}, and various approaches are possible. The main advantages offered by BNs are two:
\begin{itemize}
	\item the possible correlations among different bank processes can be captured;
	\item the information contained into both assessments and historical loss data can be merged in a natural way.
\end{itemize}

One approach \cite{adusei-poku-et-al, neil-andersen-hager} may be to design a completely different network for each bank process, trying to determine the relevant variables (in the context of each process) and the causal relationship among them; this kind of network has only one output node which typically represents the loss distribution for the process under investigation; the correlations among different processes can be captured by building a ``super-network'' which contains all the networks built for the single processes and in which the nodes representing the loss distributions of the processes may be connected by links. Since it deals with the variables governing the underlying dynamics of the bank, this approach seems to be the most convincing; nevertheless it suffers from some drawbacks as regards the practical implementation inside a bank: i) domain experts are needed for each process, in order to properly identify the variables and to define the topology of each network; ii) if the historical data need to be used, a system monitoring all the included variables with an acceptable frequency and accuracy has to be built; since this kind of network can easily reach large sizes (tens of variables), managing such systems is quite challenging and resource demanding for a mid or small sized bank.

A simpler approach \cite{cornalba-giudici_2004} is to design a unique network composed by a node for each process which represents its loss distribution; all nodes are output nodes and the operational losses are sufficient to build a historical database, so that collecting the data and managing them is much easier; in comparison with the previous approach even the experts'~task becomes simpler since their assessment reduces to an estimate of some parameters of the loss distributions; the correlations among different processes are captured through the topology of the network. This approach resembles a way of reasoning typical of the field of Complex Systems: the information carried by the ``microscopic'' degrees of freedom (the relevant variables identified in the first approach) is integrated out and the state of the system is represented by some ``macroscopic'' quantity (the loss distribution in the second approach).

Let us remark that, as regards the practical implementation inside a bank, the difference between the two approaches is huge: in the first approach tens of variables for each process need to be monitored, while in the second approach only one variable per process (the registered losses) has to; considering that an AMA-oriented bank has to track its own internal losses in any case, the cost of the proposed implementation is minimum.

\section{Bayesian Networks} \label{sec:bayesian networks}
Before defining a Bayesian Network \cite{pearl, jensen, neapolitan} let us introduce some general definitions about graphs. A directed graph is defined by a set of nodes $\{X_1, X_2, \ldots, X_N\}$ and by a set of directed links between couples of nodes; a node $X_1$ is said to be a \emph{parent} of the node $X_2$ if a directed link $X_1 \rightarrow X_2$ exists; a node $X_2$ is said to be a \emph{descendent} of the node $X_1$ if a directed path which starts at $X_1$ and ends at $X_2$ exists; if no such a path exists the node $X_2$ is said to be a \emph{non-descendent} of the node $X_1$; a directed path from a node to itself is called a \emph{directed cycle}; a directed graph containing no directed cycles is called \emph{Directed Acyclic Graph} (DAG).

In order to define a Bayesian Network two elements are necessary: a set of random variables $V=\{X_1, X_2, \ldots, X_N\}$ and a DAG whose nodes correspond to the random variables in $V$; note that, since there is no risk of ambiguity, we use the symbol $X_1$ to refer both to the random variable and to the corresponding node. Moreover, the joint \emph{Probability Distribution Function} (PDF) $P(X_1, X_2, \ldots, X_N)$ must satisfy the Markov condition, i.e.\ each random variable $X_i$ and the set of all its non-descendents must be conditionally independent, given the set of all its parents. It can be proved for discrete variables (which turns out to be our case) that the Markov condition easily allows one to calculate the joint PDF as:
\begin{equation} \label{eq:joint_PDF}
	P(X_1, X_2, \ldots, X_N) = \prod_{i=1}^{N}{P(X_i | \text{Pa}_i)},
\end{equation}
where $\text{Pa}_i$ is the set of random variables whose corresponding nodes are parents of the node $X_i$.

Both the directed links appearing in the DAG and the values of the conditional probabilities $P(X_i | \text{Pa}_i)$ can be learned from a dataset whose records hold the values assumed by each $X_i$ in \emph{independent} experiments. Even if we are not dealing here with the problem of a rigorous definition of what independent experiments are, we will be more formal about this point because it is the core of our implementation. Let us associate a random variable to each node, and to each experiment: $X_i^{(p)}$ is the random variable associated with the $i$th node and with the $p$th experiment. The $p$th and the $q$th experiments ($p\neq q$) are said to be independent if $X_i^{(p)}$ and $X_j^{(q)}$ are independent $\forall \, i$ and $j$.

The algorithm we use for learning the structure of the network is described in Ref. \cite{hugin_manual}. In the following we sketch the main steps it is founded on: i) the presence of conditional independencies among the random variables is checked performing the same test as in Tetrad II \cite{scheines-et-al}; if two variables are \emph{not} found to be conditionally independent given any set of variables, an undirected link is placed between the corresponding nodes; ii) the PC algorithm \cite{spirtes-et-al} is used to direct the links.

\section{Different-times correlations} \label{sec:different-times correlations}
One of the fundamental reasons to use BNs to estimate the OR is that if correlations do exist among different processes they can be captured through the network topology; however the correlation can extend arbitrarily over time: an example will help to clarify. Suppose that an employee violates the transaction control system with fraud purposes: he succeeds in his aim and a money loss is generated in some process of the bank. As a side effect a part of the IT infrastructure is damaged, but the failure is discovered and repaired only a week later: a loss is generated in the process of machinery servicing with a one week lag. At the same time the system remained partially unavailable and a certain amount of  transactions failed, eventually generating losses delayed up to a week in many other processes. 

In order to understand the importance of this point we need to look at the structure of a database of historical losses: each record holds the daily losses classified by the process in which they occur. The example should have made it obvious that the losses registered in different days cannot be considered as originated by independent experiments (as defined in Section \ref{sec:bayesian networks}), so a database with this structure is in principle useless for learning purposes. To overcome this limitation we propose a new approach:  the losses are averaged over a certain time interval such that the correlations of the \emph{averaged} losses vanish at different times, but are still present at the \emph{same} time. Moreover, if the losses registered in a process influence the losses of other processes in a way similar to the one illustrated in the previous example, it is natural to expect that the correlations decrease with time.

In the previous example the losses are directly correlated: this implies that the loss distributions over some time horizon are the only relevant quantities. It is possible indeed to derive the loss distributions using more sophisticated tools, e.g.\ the LDA uses both the distribution of the frequency with which a loss occurs in a certain time horizon and the distribution of the severity of the losses over the same-time horizon; in such an approach it may be natural to include correlations among the frequencies or among the severities (or correlations among frequencies and severities), but it is not so simple and immediate to take into account direct correlations among losses.

In our approach the original database is replaced by a new database (which will be called the \emph{extracted database}) of averaged losses whose number of records is $\frac{L}{T}$, being $L$ the number of records into the original database and $T$ the time interval we are averaging on. Suppose e.g.\ that the original database contains the daily losses and $L$ equals to $1$ year, and that $T=90$ is the time interval we are looking for: this means that the \emph{average} losses of a quarter of year are not correlated with the \emph{average} losses of \emph{another} quarter, but the \emph{average} losses recorded by different processes in the \emph{same} quarter are still correlated among themselves; different quarters may be considered independent experiments, thus the extracted database can be used for learning purposes.

Since we expect that the correlations decrease with time, the average over every $T > 90$ makes the different-times correlations vanish and, at least in principle, there is no reason to choose one particular value of $T$ in this interval: for this reason, to consider the whole procedure consistent, we require that the final results depend weakly on $T$.

\section{Learning Bayesian Networks by aggregate losses} \label{sec:learning BNs by aggregate losses}
In Section \ref{sec:different-times correlations} the idea of averaging the losses over a certain time interval is introduced. What we actually do is to \emph{sum} all the losses belonging to the same process and the same-time interval: the $k$th record in the extracted database contains the aggregate loss of the records from $(k-1) \, T + 1$ to $k \, T$, obviously retaining the process classification. Let us suppose again that $T=90$: the first (second, \dots) record in the extracted database contains the aggregate loss of the records from $1$ to $90$ ($91$ to $180$, \dots) in the original database. Summing is equivalent to averaging but, as we are going to see, makes much more sense in view of the VaR calculation.

After the new database has been extracted, we can start to build the network defining the nodes and the allowed states of the associated variables: we set the number of states $n$ to be equal for all the variables; the bins are equally spaced, being $0$ the lower limit and the maximum \emph{average} loss of each process the higher limit. 

The extracted database is then used to learn the structure of the network and the conditional probabilities. As hinted in Section \ref{sec:introduction}, another reason why BNs seems to be suitable for OR estimation is that they allow the integration of the information coming from the historical database with the information coming from experts'~assessment. Topology constraints can be imposed before the structure learning is performed, while \emph{a prior} knowledge can be embedded properly setting the marginal distributions of each variable before the conditional probability learning is performed. However, we are mainly interested in studying the correlations of the losses and thus we choose neither to impose topology constraints, nor to embed any prior knowledge about the marginal distributions of the variables.

The joint PDF can then be derived using \eqref{eq:joint_PDF} and the marginal PDF for each variable calculated. We recall here that the database entries are values assumed by the random variables associated with the nodes (see Section \ref{sec:bayesian networks}): if the database used for the learning procedure contains the cumulative losses of a quarter (classified by process), the marginal PDFs obtained as the output of the BN will be the loss distribution per quarter (classified by process). Let us note that, provided that $T=90$ is such that the different-times correlations vanish, it is reasonable to consider the loss distributions relative to \emph{different} quarters to be independent. Before proceeding we make the further assumption that the loss distributions per quarter are the same for each quarter; even if this assumption may sound quite strong, it is in no way different from the usual assumption which is made in the context of other AMA approaches like the LDA. In fact the LDA identifies the capital charge required to face Operational Risk in the \emph{next} year as the $99.9$ percentile of the yearly total loss distribution which is obtained using historical data relative to the \emph{past} year(s), implicitly assuming that yearly loss distribution does not depend on the year. This assumption is certainly stronger for smaller values of $T$ and, even if reasonable for $T$ equal to $1$ year, it may be not for $T$ equal to $1$ quarter: for this reason we test our algorithm through a range of values of $T$ whose maximum ($T=240$) is comparable to $1$ year.

The loss distributions per year can be calculated by convoluting the loss distributions per quarter $4$ times by itself; an explanation follows. Let $P_i^{\mathcal{T}}$ be the random variable representing the losses over some time horizon $\mathcal{T}$ and $p_i^{\mathcal{T}}$ the relative distribution, so that $P_i^{360}$ is the random variable representing the loss distribution per year and $p_i^{360}$ its distribution; moreover, let $P_i^{90 \,(q)}$ be the random variable representing the loss of the $i$th process over the $q$th quarter and $p_i^{90 \,(q)}$ its distribution; clearly one has that
\begin{equation*}
	P_i^{360} =\sum_{q=1}^{4}{ P_i^{90 \,(q)} }.
\end{equation*}
Since we supposed that for $T=90$ the different-times correlations vanish, $P_i^{90 \,(q)}$ for $q=1, \ldots, 4$ are independent variables and thus the distribution of their sum is equal to:
\begin{equation*}
	p_i^{360} = p_i^{90 \,(1)} * \; p_i^{90 \,(2)} * \; p_i^{90 \,(3)} * \; p_i^{90 \,(4)},
\end{equation*}
where $*$ stands for the convolution product. Since we deal with discrete distributions, the convolution product of $p_i$ by $q_i$ is defined as:
\begin{equation*}
		\left( p_i * q_i \right) (k) = \sum_{m=\max(1, k+1-n)}^{\min(k,n)} {p_i(m) q_i(k - m + 1)},
\end{equation*}
where $p_i(k)$ is the value of $p_i$ in the $k$th bin. Assuming that the loss distributions per quarter are the same for each quarter, i.e.\ $p_i^{90 \,(q)} = p_i^{90}$, for $q=1, \ldots, 4$, we have:
\begin{equation*}
	p_i^{360} = p_i^{90} * \; p_i^{90} * \; p_i^{90} * \; p_i^{90}.
\end{equation*}

In order to compare the results obtained for different values of $T$, we calculate the loss distributions and the VaR over a fixed time horizon and with a confidence level of $99.9\%$; since the original database we use is artificial (see Section \ref{sec:synthetic data}), the only characteristic time length available, apart from $T$, is the length $L$ of the database (the number of time steps it contains), which seems the most natural time horizon to fix. To this purpose the previous discussion can be generalized to the calculation of $p_i^{\mathcal{T}}$, with arbitrary $T$ and $\mathcal{T} = L$. Making assumptions analogous to those made in the previous discussion we have that $p_i^L$ is obtained convoluting $\frac{L}{T}$ times $p_i^T$ by itself:
\begin{equation*}
		p_i^{L} = p_i^T \; \underbrace{* \quad \dots \quad *}_{\frac{L}{T} \; \text{times}} \; p_i^T.
\end{equation*}

Let us generalize the definition of VaR given in Section \ref{sec:introduction}: we define $\text{VaR}_i^{99.9\%, \, L}$ to be the VaR of the $i$th process over the time horizon $L$ and with a confidence level of $99.9\%$, i.e.\ the $99.9$ percentile of $p_i^L$, the loss distribution over the time horizon $L$; this means that the probability that a loss is less than or equal to the value of $\text{VaR}_i^{99.9\%, \, L}$ is registered in the $i$th process  during the time horizon $L$ is equal to $0.999$:
\begin{equation*}
	\text{Pr}[P_i^L \leq \text{VaR}_i^{99.9\%, \, L} ] = 0.999.
\end{equation*}
The total VaR is simply the sum of the VaRs of the single processes:
\begin{equation*}
	\text{VaR}^{99.9\%, \, L} = \sum_{i=1}^N {\text{VaR}_i^{99.9\%, \, L} }.
\end{equation*}

The $99.9$ percentile of the convoluted distribution (for each process) can be numerically determined in the following way: the convoluted distribution is sampled $10^3$ times and the sample is arranged in increasing order: the second largest value is the $99.9$ percentile of the convoluted distribution. Since this procedure involves sampling, it is repeated several times and the mean of the obtained $99.9$ percentiles is calculated.

As hinted before, the VaR may be calculated over every desired time horizon $\mathcal{T}$ tuning the number of convolutions, so that the $p_i^{\mathcal{T}}$s are obtained; in particular the time horizon can be set to $1$ year, as required by Basel II, performing $\frac{360}{T}$ convolutions, if each record in the original database contains the daily losses.

\section{Synthetic Data} \label{sec:synthetic data}
In order to investigate our approach, we developed  a reliable and tunable database of  synthetic internal losses: in this way we are able to control the correlations between the different processes and some inherent features of each process. 

We consider the historical losses of each process as a time series and, inspired by Ref. \cite{sto_tsg}, generalize a stochastic algorithm for generating a set of time series. We point out that this procedure allows us to impose, at least in principle, arbitrary cross-correlation functions between each pair of generated time series, as well as the auto-correlation function and distribution for each generated time series. The algorithm defines a score to be assigned to a set of time series and generates a new set of time series such that its score cannot be lower than the one assigned to the previously generated set; the mechanism for generating the new set of time series is to randomly exchange two values belonging to one of the time series of the previously generated set.

The steps of the algorithm are the following: i) for each process, $L$ values are drawn from an arbitrary distribution; the order in which the values are extracted is considered to be a temporal order, so let us call the extract values $l_i(s)$, where the subscript $i=1,\dots,N$ indexes the process and the argument $s=1,\dots,L$ defines the temporal ordering. ii) The following quantity is calculated:
\begin{equation} \label{eq:ls_distance}
	D = \sum_{i,j=1}^{N}\sum_{t=1}^{L-1}{\left[ c_{ij}(t) - C_{ij}(t) \right]^2 },
\end{equation}
where $N$ is the number of processes, $C_{ij}$ are the imposed cross-correlation (or auto-correlation) functions, while $c_{ij}$ are the cross-correlation (or auto-correlation if $i=j$) functions calculated from the generated data:
\begin{equation} \label{eq:c}
	c_{ij}(t) =  \frac{1}{\cov(l_i,l_j)} \left[ \frac{1}{L-t} \sum_{s\leq L-t}{l_i(s) l_j(s+t)} - \langle l_i \rangle \langle l_j \rangle \right],
\end{equation}
with $\langle l_i \rangle = \frac{1}{L} \sum_{s\leq L}{l_i(s)}$ and $\cov(l_i,l_j) = \left\langle \bigl( l_i - \langle l_i \rangle \bigr) \bigl( l_j - \langle l_j \rangle \bigr) \right\rangle$. From \eqref{eq:c} it follows that $c_{ij}(0)=1$: in other words, because of its normalization, $c_{ij}$ carries no information about the same-time correlations; in order to make the whole procedure consistent $C_{ij}(0)$ must also be equal to $1$: this explains why the summation over $t$ in \eqref{eq:ls_distance} starts from $1$ and not from $0$. $D$ plays the role of the inverse of a score assigned to the generated time series: the larger the value of $D$, the farther the correlation functions calculated from the generated data are from the imposed correlation functions. iii) A new set of time series is generated by exchanging at random two values belonging to a randomly selected series; then $D$ is recalculated for the new set of generated time series; iv) if $D$ has decreased, the exchange between the two values performed in the step (iii) is accepted, otherwise it is rejected: clearly the value of $D$ associated with the new set of time series is lower than or equal to the value of $D$ associated with the previously generated set of time series. The algorithm is iterated so that the value of $D$ progressively decreases or, equivalently, the score associated with the set of time series increases. As $c_{ij}$ are not limited, $D$ cannot be normalized and thus a threshold for $D$ below which the algorithm is halted cannot be set. We rather choose to iterate the algorithm until $D$ reaches a plateaux. 

Since we are interested in the change of the correlation between different processes with respect to the time interval $T$ over which the losses are averaged, we imposed auto-correlation and cross-correlation functions of the form:
\begin{equation} \label{eq:C}
	C_{ij}(t) = e^{- \frac{t}{\tau_{ij}}},
\end{equation}
in fact making such a choice implies that the different-times correlation between the processes $i$ and $j$ should be significantly reduced averaging over a time interval $T \simeq \tau_{ij}$. 

Even though the algorithm allows us to impose both distributions and $C_{ij}$, in practice a certain degree of compatibility may exist between them: this means that, even if \eqref{eq:ls_distance} reaches a plateaux, still $c_{ij}$ and $C_{ij}$ are significantly different. In order to overcome this limitation the algorithm is slightly modified in the following way: we generate series which are indeed longer than $L$ so that a larger basin of values that may fulfill the imposed constraints is available; e.g. suppose that the values of the series are drawn from a uniform distribution and that the imposed $C_{ij}$ have an higher degree of compatibility with another distribution: a subset of values belonging to this distribution will be selected by the algorithm. The modified algorithm obviously alters the imposed distributions; however we see no reasons to impose strict constraints on the distributions and, on the other hand, as we are interested in studying the correlations between the processes, we need a high accuracy in reproducing the $C_{ij}$.

\section{Results} \label{sec:results}
We investigate a sort of \emph{toy model} whose number of processes is limited to $N=3$; this choice is the result of a trade-off between our need to consider a system complex enough to have a reasonable number of correlated processes and the convenience of using series long enough to be able to carry out the average over time and still have a sufficient number of data to perform the learning of the network. With $L=5000$ it is possible to average over $240$ steps and still have $20$ patterns left for the learning. The number of states for each node has been set to $n=5$ based on computational needs as well: in fact the number of states of the entire system scales exponentially with $n$ and, given a number of patterns available, the learning procedure is more accurate for a smaller number of states of the system.

\enlargethispage*{\baselineskip}

\begin{figure}[t]
	\centering
	\includegraphics[width=1 \columnwidth]{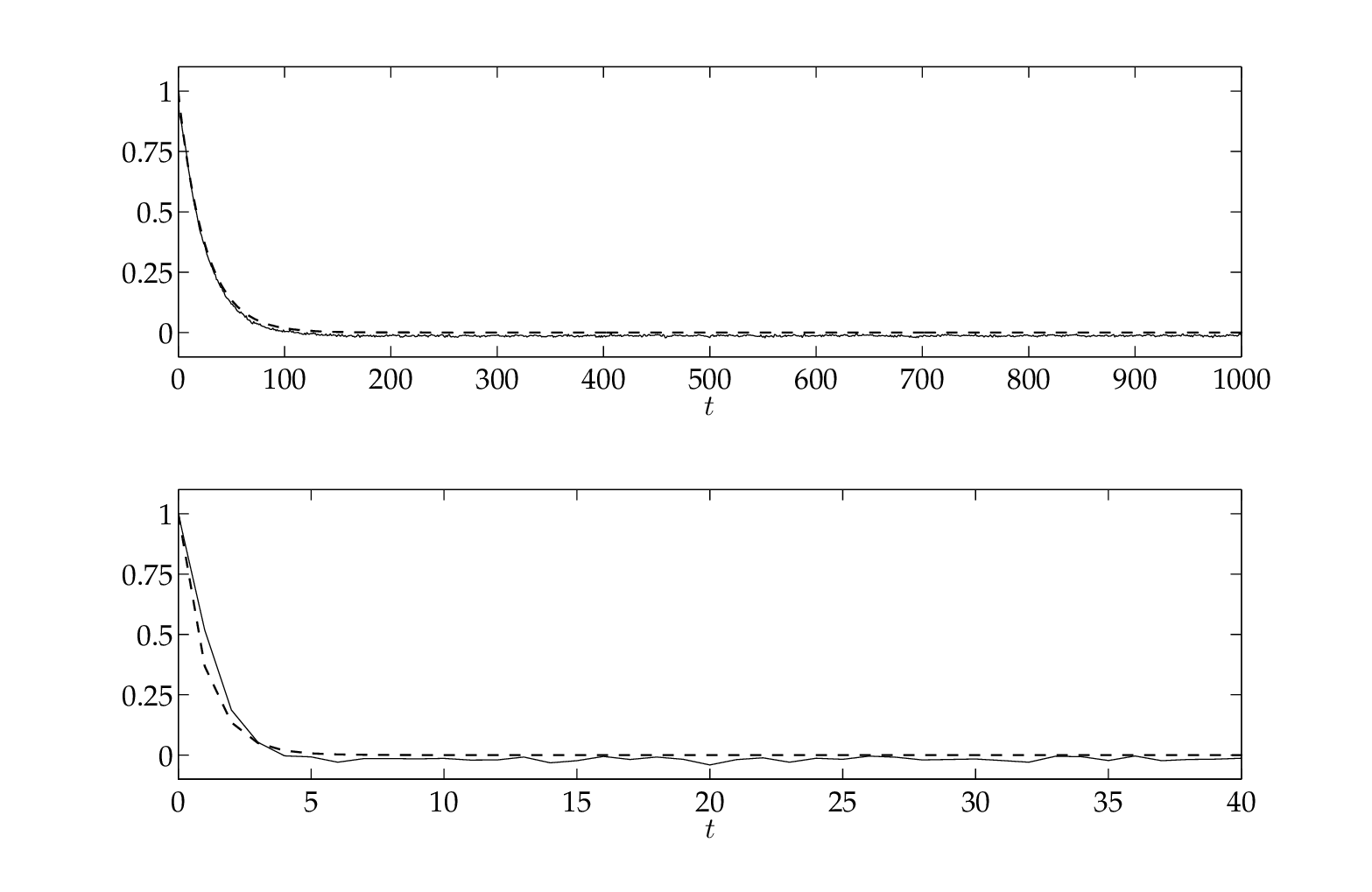}
	\caption{Top panel: imposed cross-correlation function $C_{12}$ (dashed line) and obtained cross-correlation function $c_{12}$ (solid line), with no average on time; for the sake of readability only the first $1000$ values are shown. Bottom panel: imposed cross-correlation function $C_{12}$ with scaled decay time $\frac{\tau}{25}$ (dashed line) and obtained cross-correlation function $c_{12}$ (solid line) averaged over a time interval $T=~25$; for the sake of readability only the first $40$ values are shown.}
	\label{fig:corr_functions}
\end{figure}

The negative exponential distribution has shown to be compatible with \eqref{eq:C} if the decay matrix is homogeneous, i.e.\ $\tau_{ij} = \tau, \;\; \forall \; i$ and $j$ and if $\frac{\tau}{L}$ is not too large. In the top panel of Fig.\ref{fig:corr_functions} both $c_{ij}$ and $C_{ij}$ are shown  for $\tau=25$. In order to simulate different kinds of processes their means have been set respectively at $100$, $50$ and $10$. Using a larger basin of values as described in Section \ref{sec:synthetic data} both the mean and the variance of the distributions do not significantly change, but a heavier tail appears.

As is shown in the bottom panel of Fig.\ref{fig:corr_functions}, averaging over a time interval $T$ leaves the form \eqref{eq:C} unchanged with a new decay time equal to $\frac{\tau}{T}$. This actually means that, at the cost of reducing the length of the time series, averaging effectively removes the different-times correlations: in particular when $T = 2.4 \, \tau$ ($T \simeq 60$) all the different-times correlations are reduced to $0.1$ and for $T \ge 80$ they can be considered effectively extinguished.

Since $C_{ij}$ carry no information about the same-time correlations (see Section \ref{sec:synthetic data}), in order to study them we look at the learned structure of the networks: in Tab.\ref{tab:nets} it is shown that the number of links decreases as $T$ increases. This is somewhat expected since, as $T$ increases, the size of the extracted database reduces and it becomes more and more difficult to learn from it. However for $T \ge 80$ the algorithm of structure learning still detects the presence of some links: since the different-times correlations are extinguished for such a large $T$, they must be due to the survived same-time correlations.

In order to evaluate the consistency of the whole procedure we require that, for values of $T$ such that the different-times correlations can be neglected, the value of $\text{VaR}^{99.9\%, \, L}$ depends weakly on $T$.

\begin{figure}[t]
	\centering
	\includegraphics[width=1 \columnwidth]{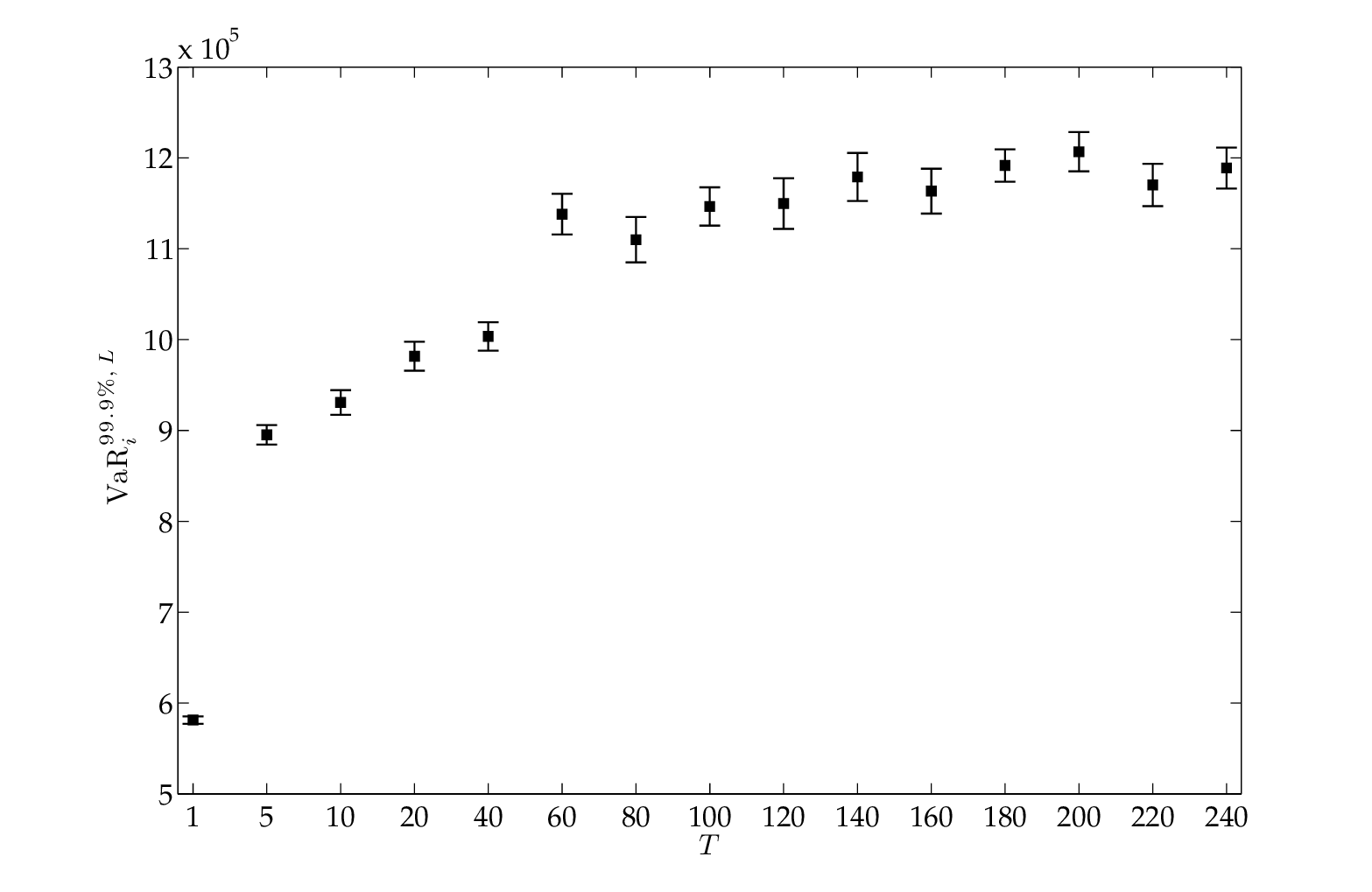}
	\caption{$\text{VaR}^{99.9\%, \, L}$ with respect to the time interval $T$ over which the average of the losses is performed; each point is the mean over $30$ realizations of the procedure described in Section \ref{sec:learning BNs by aggregate losses} and the error bars span over one standard deviation. For $T \ge 60$ the values of $\text{VaR}^{99.9\%, \, L}$ are compatible among themselves. For $T<60$ the values are not reliable because the records in the extracted database cannot be considered independent.}
	\label{fig:VaR}
\end{figure}

In Fig.\ref{fig:VaR} the values of $\text{VaR}^{99.9\%, \, L}$ with respect to $T$ are represented; each point is the mean over $30$ realizations of the procedure described in Section \ref{sec:learning BNs by aggregate losses} and the standard deviations are also shown. Indeed from Fig.\ref{fig:VaR} it can be seen that for $T \ge 60$ the values of $\text{VaR}^{99.9\%, \, L}$ are compatible among themselves. On the other hand, for $T < 60$ the different-times correlations are still present and the records belonging to the extracted database cannot be considered independent; nevertheless the learning algorithm for BNs considers them to be independent (see Section \ref{sec:bayesian networks}) and returns unreliable loss distributions: the corresponding values of $\text{VaR}^{99.9\%, \, L}$ are consequently also unreliable.

\begin{table}
	\centering
	\begin{tabular}{CF|CF|CF}
		\toprule
		T	&		&	T	&		&	T	&		\\
		\midrule
		
		1	&	\includegraphics[width=0.2 \columnwidth]{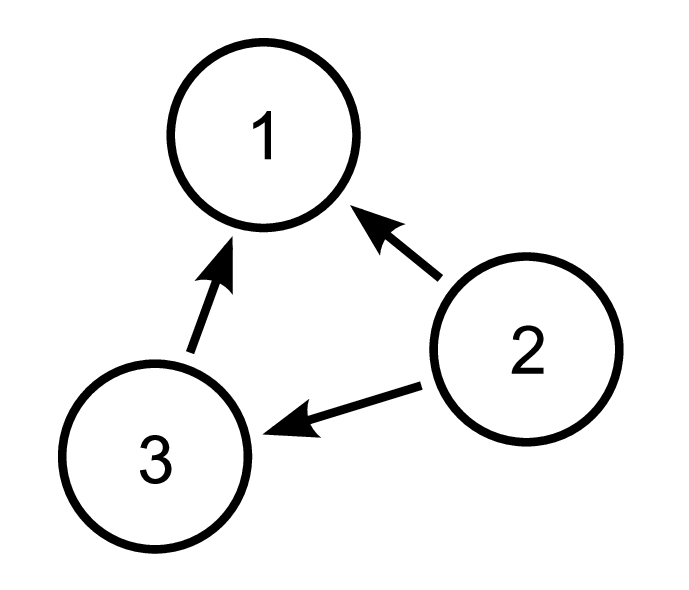}		&			60	&	\includegraphics[width=0.2 \columnwidth]{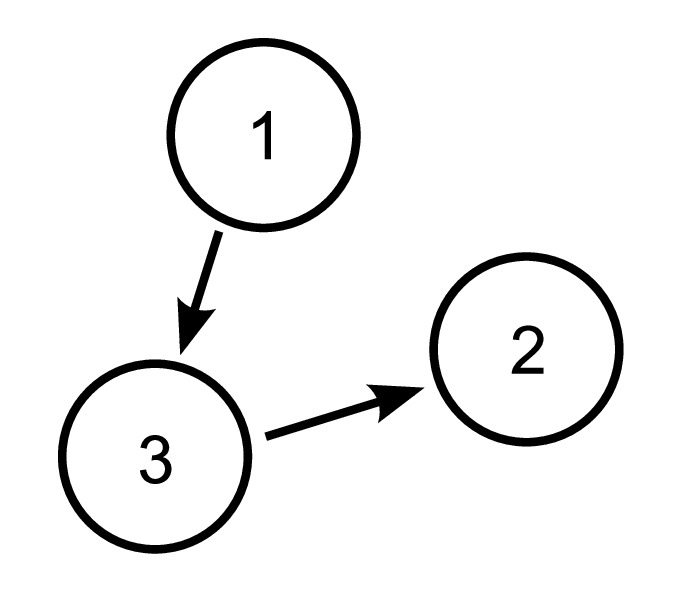}		&			160	&	\includegraphics[width=0.2 \columnwidth]{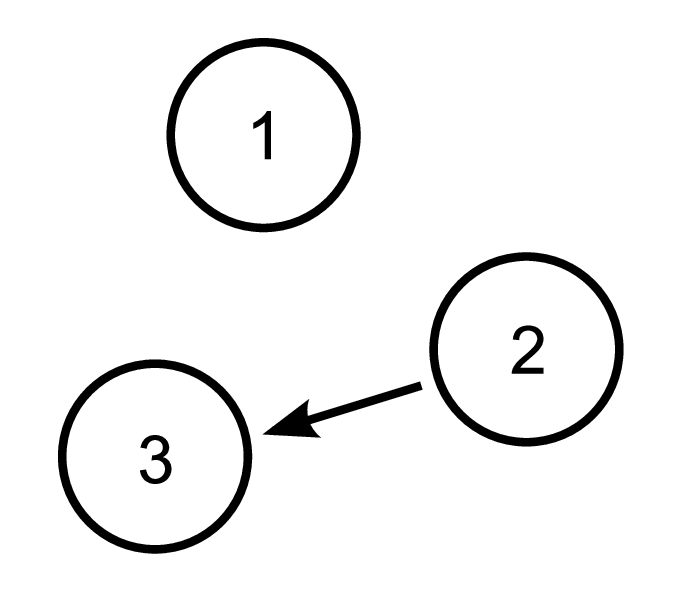}		\\
		5	&	\includegraphics[width=0.2 \columnwidth]{tab_fig1}		&			80	&	\includegraphics[width=0.2 \columnwidth]{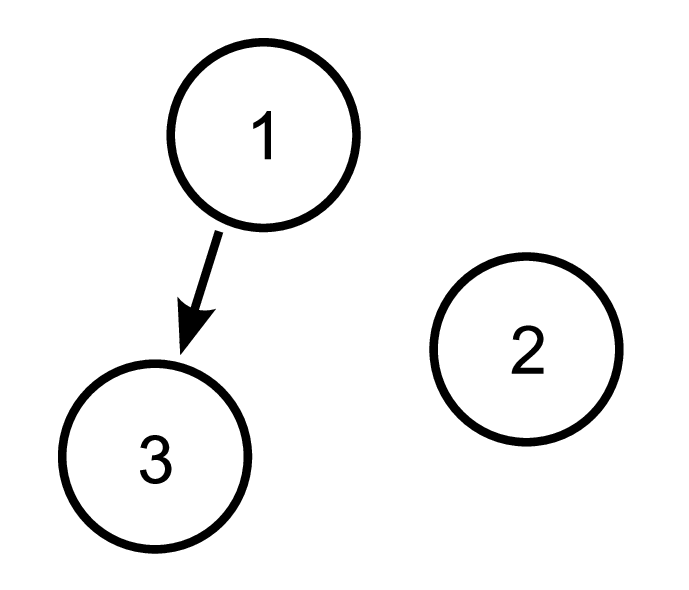}		&			180	&	\includegraphics[width=0.2 \columnwidth]{tab_fig3}		\\
		10	&	\includegraphics[width=0.2 \columnwidth]{tab_fig1}		&			100	&	\includegraphics[width=0.2 \columnwidth]{tab_fig3}		&			200	&	\includegraphics[width=0.2 \columnwidth]{tab_fig3}		\\
		20	&	\includegraphics[width=0.2 \columnwidth]{tab_fig1}		&			120	&	\includegraphics[width=0.2 \columnwidth]{tab_fig3}		&			220	&	\includegraphics[width=0.2 \columnwidth]{tab_fig3}		\\
		40	&	\includegraphics[width=0.2 \columnwidth]{tab_fig2}		&			140	&	\includegraphics[width=0.2 \columnwidth]{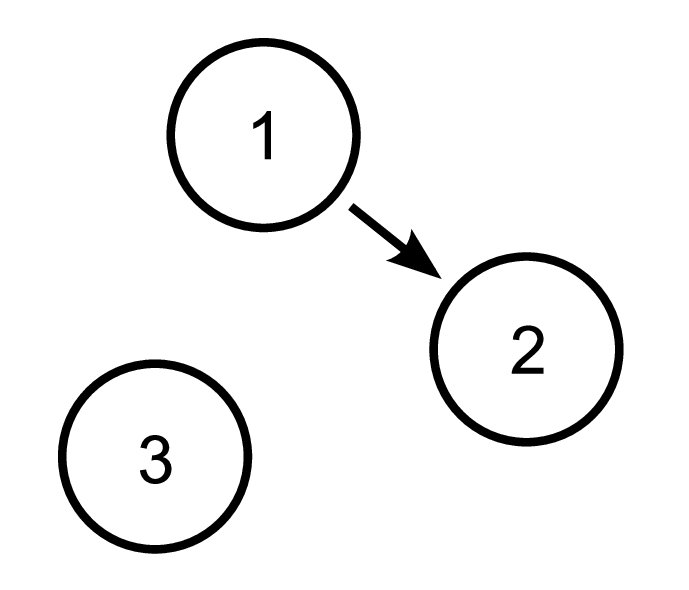}		&			240	&	\includegraphics[width=0.2 \columnwidth]{tab_fig3}		\\
		
		\bottomrule
	\end{tabular}
	\caption{The topology of BNs as the time interval $T$ over which the losses are averaged varies. The links are more difficult to detect as $T$ increases because the size of the extracted database used for the structure learning reduces. For $T \simeq 60$ the different-times correlations are reduced to $0.1$, while for $T \ge 80$ they are extinguished and only the same-time correlations remain (see Section \ref{sec:results}).}
	\label{tab:nets}
\end{table}

\section{Conclusion}
A novel approach, based on Bayesian Networks, has been proposed for the quantitative management of Operational Risk in the framework of The New Basel Capital Accord. The main advantage of every BNs approach over the other AMAs (like the widely used LDA) is the possibility of capturing the correlations among different bank processes; however, as shown in Section \ref{sec:different-times correlations}, the different-times correlations play a significant role and are in no way negligible with respect to the same-time correlations, but (at least to the best of our knowledge) there is no other approach taking them into account. The need to deal with different-times correlations leads us to propose a solution for the problem of learning a BN using a time ordered set of operational losses (see Section \ref{sec:different-times correlations} and \ref{sec:learning BNs by aggregate losses}).

The proposed approach is validated by means of synthetic data for two reasons; the first one is methodological: in such a way it is possible to generate data containing the information that the algorithm should extract; the second reason is practical: regulatory laws on Operational Risk exist from a relatively short period (see Section \ref{sec:introduction}) and it is very difficult to obtain ``experimental'' data on operational losses which are accurate and reasonable in size.

At the moment the number of processes is limited to $N=3$ for computational reasons: the stochastic algorithm described in Section \ref{sec:synthetic data} has to explore a configuration space of size $L! \, N^2$. On one hand, keeping $L$ low would imply a lower accuracy of the learning procedure (recall from Section \ref{sec:different-times correlations} that the number of patterns used for the learning are $L/T$) and this is not acceptable from the point of view of the validation of the proposed approach; this is especially true for the higher values of $T$, those for which the approach proves to be consistent. On the other hand, since there is no assumption in our algorithm about the value of $N$, the obtained results should not depend on it, at least qualitatively.

The principal features of the proposed approach are the following: 1) the whole topology of the network is derived from data of operational losses; each node in the network corresponds to a bank process and the links between the nodes, which are drawn learning from data, model the causal relationships between the processes; this scheme seems more flexible than the classification in $56$ pairs $(\text{Business Line}, \text{Loss Event Type})$ prescribed by Basel II and has the advantage of representing both the units that generate operational losses and the relationships between them; however it has to be pointed out that the decision not to use the assessment of domain experts is motivated by the need to carefully study the correlations present in the historical data: as hinted in Section \ref{sec:introduction}, one of the possible developments is to take the assessments into account. 2) For the first time a Bayesian Network is used to represent the influence between correlated operational losses that take place in different days exploiting a dataset whose records represent losses occurred over $T$ days: using such a dataset the nodes in the network represent the aggregate loss over $T$ and the VaR over a time horizon $T$ can be computed. The extension to the VaR over the time horizon $L$ requires an additional assumption (see Section \ref{sec:learning BNs by aggregate losses}) and is performed by convoluting the probability density functions $\frac{L}{T}$ times and extracting the $99.9$ percentile of the convoluted distribution. 3) The proposed approach is tailored for a practical implementation inside a mid or small sized bank: since the network contains only nodes representing the loss distributions over some time horizon, only the losses occurring in the different processes have to be monitored.

\section*{Acknowledgements}
One of us (M. Bardoscia) would like to thank M.V. Carlucci for the countless suggestions and useful discussions.



\end{document}